\documentclass[eprint,amsmath,amssymb,aps,nofootinbib]{revtex4-1}

\newcommand\simlt{\lower.5ex\hbox{$\; \buildrel < \over \sim \;$}}
\newcommand\simgt{\lower.5ex\hbox{$\; \buildrel > \over \sim \;$}}

\DeclareMathAlphabet\mbi{OML}{cmm}{b}{it}

\usepackage{graphicx}% Include figure files
\usepackage{dcolumn}% Align table columns on decimal point
\usepackage{bm}% bold math

\begin{document}

\preprint{APS/123-QED}

\title{Magnetic collimation of meridional-self-similar GRMHD flows}

\author{Noemie Globus}
\affiliation{School of Physics \& Astronomy, Tel Aviv University, Tel Aviv 69978, Israel}
\author{Christophe Sauty,\\ V\'eronique Cayatte,\\ Ludwik M. Celnikier}
\affiliation{LUTH, Observatoire de Paris, CNRS \& Universit\'e Paris Diderot, 92190 Meudon, France} 

\begin{abstract}
We present a model for the spine of relativistic MHD outflows in the Kerr geometry. Meridional self-similarity is invoked to derive semi-analytical solutions close to the polar axis. The study of the energy conservation along a particular field line gives a simple criterion for the collimation of jets. Such parameter have already been derived in the classical case by Sauty et al. 1999 and also extended to the Schwarzschild metric by Meliani et al. 2006. We generalize the same study to the Kerr metric. We show that the rotation of the black hole increases the magnetic self-confinement. 
\end{abstract}

%\keywords{}   

\maketitle

\date{\today}% It is always \today, today,
             
\section{Introduction}             

Several studies have contributed to show the importance of black hole rotation in AGN jet formation. AGN jet classification mainly relies on orientation 
effects and relativistic doppler boosting. However, they cannot explain neither the dichotomy between radio loud and radio quiet AGN, nor the difference 
between FRI and FRII jets. There are two main theories to interpret the different characteristics of radio loud and radio quiet galaxies. 
The morphological differences may be explained by the different physical properties of the environment in which the relativistic jet propagates \citep{DeYoung93, Bicknell95, Laingetal99, Gopal-KrishnaWiita00,Melianietal08}.
The dichotomy can also be explained by involving a difference in the nature of the central engine,  the spin of the central black hole, the accretion rate and the jet composition \citep{Baumetal95, Reynoldsetal96, Meier99, MelianiKeppens09, McKinneyBlandford09}. The discovery of two sub-classes of FRII galaxies does not allow to solve the problem of the dichotomy. The Hybrid Morphology Radio Sources have two radio lobes which exhibit a different FR morphology and can not be explained without external medium or jet power differences between the two sides of the host galaxy \citep{Ceglowskietal13}. In Double-Double Radio Galaxies multiple pairs of lobes are seen and are interpreted as different episodes of jet activity \citep{Brocksoppetal11,KonarHardcastle13} leading to the possibility that jet interruption occurs in radio galaxies.
Finally, a combination of external and engine factors has to be invoked to explain the FRI/FRII dichotomy, as we have suggested in \citep{Melianietal10}.
However, in this paper we could not study the effect of the black hole spin because the analytical model was based on a Schwarzschild metric. 

The dichotomy between radio loud and radio quiet sources has a result of the spin of the black hole has been explored analytically and numerically 
\citep{Tchekhovskoy10}.  Moreoever \citet{NarayanMcClintock12} have recently demonstrated that there is a strong correlation between the power of the 
jet and the spin of the black hole. The spin of the black hole may also explain the high precession that is observed in some jets 
\citep{Begelman80}. In fact, as already suggested by \citet{BlandfordZnajek77}, the rotational energy of the central black hole is a tremendous spring for
energy invoked to explain emission of plasma flows. They were the first to propose a magnetospheric model in the force free limit as a 
source for extracting rotational energy from the black hole. 
This model has been extensively discussed in the frame of ideal MHD \citep{Phinney83, Camenzind86, Takahashi90, 
Okamoto92, Fendt97, BeskinKuznetsova00}. The energy extracted under the form of Poynting flux depends on the
rotational speed of the magnetic fieldlines, the spin of the black hole, and as shown recently by \citep{GlobusLevinson13}, on plasma injection on magnetic field lines. Other 1D models allowed to study force free magnetospheres around black holes, first in Schwarzschild metrics by \citep{Uzdensky04} and in Kerr metrics by \citep{Uzdensky05}, with a more complexe 
magnetic configuration and taking into account the connexion between the magnetosphere and the accretion disk. This last study shows that the higher the black hole spin the smaller the magnetic dead zone of the 
magnetosphere. S. Komissarov was the first to perform numerical simulations of the Blandford Znajek scenario \citep{Komissarov01,Komissarov04a,Komissarov04b,Komissarov05}.

\begin{figure}
\includegraphics[width=6.cm]{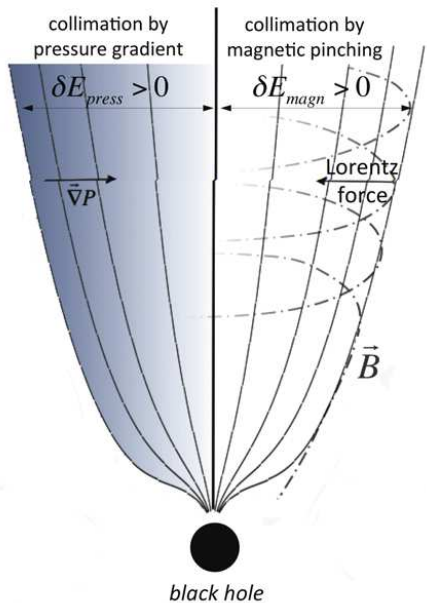}
\caption{\label{fig1}Illustration of the self-collimation processes.}
\end{figure}

Besides the strong acceleration that they undergo, AGN outflows appear to be highly collimated. The question we want to address is whether 
there is a correlation between the collimation of the jets and the black hole spin. We propose a model for a rotating black hole based on our previous 
steady axisymmetric analytical model \cite{Melianietal10}. 
The problem of collimation in all its complexity must include the interaction between the jet and the external 
medium (see e.g. Levinson \& Begelman \citep{LevinsonBegelman13}, and references therein). In the following, we focus on the self-collimation processes where it is the combination of gas pressure and magnetic fields that acts to confine the flow, as illustrated in Fig.\ref{fig1}. Magnetic-self confinement requiere an ordered magnetic field anchored onto the black hole or the inner disk region. The poloidal component of the Lorentz force has a collimating effect \cite{ChanHenriksen80}. Therefore, we need to measure separately the contribution of the gas pressure and magnetic fields. In the 
framework of the $\theta$-self-similar model of \cite{ST94}, we solve the GRMHD equations in the Kerr background. In this 
model, two parameters give the magnetic and pressure confinement, as explained in \cite{STT99}. This model has been successfully applied to model 
AGN spine jets \cite{Melianietal10} and the criterion for magnetic collimation derived in the case of a non rotating black hole. In this paper we investigate the 
effect of the black hole spin on the efficiency of the magnetic confinement. 

The paper is organized as follows.  The model is introduced in Sect.\ref{model-des}. Using the $3+1$ formalism \citep{Thorne86} we present the general equations that describe an 
axisymmetric, stationary, ideal MHD flow in the gravitational potential of a Kerr black hole. The complete set of 3 + 1 equations is given by
\citet{MobarryLovelace86} in Schwarzschild geometry. We compile in Sect.\ref{equations} the complete set in Kerr geometry and obtain an original formulation of the energy and momentum equations. In Sect.\ref{model} we present the mathematical formalism and the assumptions leading to the self-similar model, an extension of the non relativistic meridionnaly self-
similar model \citep{ST94} to the case of relativistic jets around rotating black holes. In Sect.\ref{results} we present  the collimation 
criterion established by \citet{STT99} extended to this metric and show that the rotation of the black hole induces a more efficient 
magnetic collimation of the jet.

\section{A self-similar model for ideal relativistic MHD jets}\label{model-des}

The common picture for the structure of relativitic jets is the two-component model  \citep{Sol89},
where a relativistic $e^{+}e^{-}$ plasma is accelerated in the polar region of the central black hole and is surrounded by a baryonic component coming 
from the accretion disk. In AGN, pair injection in the black hole magnetosphere may arise from annihilation of MeV photons if the 
accretion rates are sufficient, or from pair cascades induced by a potential drop in charge-starved regions \citep{LevinsonRieger11}. 
When the jet is emitted in the region very close to the central engine, frame-dragging effects may play an important role on 
self-collimation processes. To adress this question we therefore need to model the inner part of the jet close to the polar axis.

The dynamics and the geometry of a magnetized, relativistic plasma flow around a rotating black hole are described by exact solutions to the general relativistic MHD equations in the Kerr metric. A standard treatment to reduce the stationnary and axisymmetric GRMHD equations to a system of ordinary differential equations is to adopt a self-similar geometry \textit{i.e.} to make the assumption of a scaling law of one of the variable as function of one of the coordinates (see \citet{VlahakisTsinganos98} for a general classification of self-similar models). Analytical solutions using radial self-similarity have been derived from GRMHD equations in order to model disk winds \citep
{VlahakisKonigl03a, VlahakisKonigl03b, VlahakisKonigl04}. However those models fail to describe the regions close to the 
 rotational axis. A meridional self-similar treatment (i.e., similar in the $\theta$-direction) is necessary to model the spine
jet where the outflow is rather driven by the thermal pressure. In those models, the $\theta$-dependence is prescribed \textit{a priori} while the radial dependence is derived from the MHD equations.  This modeling does not require the use of a polytropic equation of state. The local equation of state can be derived a posteriori from the calculated solutions.

For our investigation, we chose the analytical model for stellar jets developed by \citet{ST94}. In this model, outflow solutions are super-Alfv\'enic, and one class of the solutions provides self-confined jets \citep{STT99}.
This model has been extended to the general relativistic case in the Schwarzschild geometry \citep{Meliani06} and applied to AGN jets. 
The collimation of the relativistic solutions results from the distribution of the total electromagnetic energy across the jet,
as compared to the corresponding distribution of the thermal and gravitational energies. 
The FRI/FRII dichotomy was explained by the magnetic rotator efficiency \citep{Melianietal10}. 
In this paper, we extend the model to the Kerr geometry to study the effect of frame-dragging on  jet collimation. 
The Kerr metric is axisymmetric and hence adapted to this assumption. 
In the following, we make a Taylor expansion of all physical quantities with small $\theta$. 
Such a treatment allows to study the physical properties of the outflow close to its rotational axis but does not imply any 
restriction on the black hole rotation going from zero to maximum value.

\section{Basic equations}\label{equations}
\subsection{Flow equations} 
In the following we use the 3+1 decomposition of the Kerr spacetime to write the GRMHD equations. The rotating black hole is described by two parameters, $m={GM}/{c^2}$ and $a={J}/{Mc}$, where $M$ is the mass of the black hole and $J$ its specific angular momentum, respectively.  The Kerr metric writes,   
\begin{align}
ds^{2} =-h^2c^{2}dt^{2}+h_1dr^{2}+h_2^2d\theta^{2}+\varpi^2 (d\varphi-\omega dt)^{2}\,,
\end{align}
where $h={\rho \sqrt{\Delta}}/{\Sigma}$ is the lapse function and $\omega={2mrac}/{\Sigma^{2}}$ the angular velocity of the zero angular momentum observer (hereafter ZAMO) as seen from infinity, with $\Delta=r^{2}+a^{2}-2mr$, $\rho^{2}=r^{2}+a^{2}\cos^{2}\theta$, $\Sigma^{2}=(r^{2}+a^{2})^
{2}-a^{2}\Delta \sin^{2}\theta$.

The 4-velocity of a particle can be decomposed as 
$u^{a}=\gamma\left(c\,\vec{e}_{\hat{t}}+V_{r}\,\vec{e}_{\hat{r}}+V_{\theta}\,\vec{e}_{\hat{\theta}}+V_{\varphi}\,\vec{e}_{\hat{\varphi}}
\right)$
where $(V_{r},V_{\theta},V_{\varphi})$ are the components of the 3-velocity $\vec{V}$ relative to the ZAMO, and $\gamma=(1-V^2/
c^2)^{-1/2}$ is the Lorentz factor.
In the following, we use CGS units. The letters $i$,$j$,$k$... represent indices in absolute space and run from 1 to 3 while $a$,$b$,
$c$... represent indices in 4-dimensional spacetime and run from 0 to 3. \\

The stress energy tensor is $T^{ab}=T^{ab}_{\textrm{hyd}}+T^{ab}_{\textrm{em}}$. 
The electromagnetic stress-energy tensor $T^{ab}_{\textrm{em}}$,
\begin{align}
& T^{ab}_{\textrm{em}} = \frac{1}{4\pi}\left( F^{ac}F^{b}_{c}-\frac{1}{4}g^{ab}F^{cd}F_{cd} \right) \label{tem}
\end{align}
satisfies to the Maxwell equations,
\begin{align}
& \nabla_{a} F^{ab}=4 \pi j^{b}\,, \label{tmaxs}\\
& \nabla_{[a}F_{bc]}=0 \,.\label{tmax}
\end{align}
The electric current density 4-vector $j^{a}$ is only defined by Eq. \ref{tmaxs}. We assume that the plasma is infinitly conducting. The 
electric field is null in the comoving frame, 
\begin{align}
& F_{ab}u^{b}=0 \label{frozen}\,,
\end{align}
and thus the magnetic field is frozen to the plasma,
\begin{align}
& \nabla_{a} (u^{[a}B^{b]})=0\label{tgel}\,.
\end{align}

In the 3+1 form the Maxwell equations (\ref{tmaxs}) and (\ref{tmax}) writes \citep{Thorne86},
 \begin{align}
&\vec{\nabla}.\vec{B} = 0\,,\label{flux}\\
&\vec{\nabla}.\vec{E} = 4 \pi \hat{q}\,,\\
&\vec{\nabla} \times (h\vec{E}) = \left(\vec{B}.\vec{\nabla} \frac{\omega}{c}\right)\varpi \vec{e}_{\hat{\varphi}} \label{induction}\,,\\
&\vec{\nabla} \times (h\vec{B}) = \frac{4\pi}{c}h\vec{j} - \left(\vec{E}.\vec{\nabla} \frac{\omega}{c}\right)\varpi \vec{e}_{\hat{\varphi}}
\label{ampere}\,.
 \end{align}

where $\vec{E},\,\vec{B}$ are, respectively, the electric and magnetic fields measured by a ZAMO. The Ohm's law (\ref
{frozen}) rewrites,
 \begin{align}
&\vec{E} + \frac{\vec{V}}{c} \times \vec{B}= 0\,.\label{ohm}
 \end{align}

$T^{ab}_{\textrm{hyd}}$ is the energy-momentum tensor for a perfect  fluid, 
\begin{align}
& T^{ab}_{\textrm{hyd}}=\frac{nw}{c^{2}} u^{a} u^{b} + P g^{ab} 
\end{align}
where $n$  is the proper particle number density, $w=(e+{P})/{n}$ the specific enthalpy per particle, $e$ the internal energy density 
(including rest-mass energy per particle) and $P$ the isotropic pressure, that is the sum of the kinetic pressure and the pressure associated with the MHD waves.

The particles constituting the jet plasma may have two origins. The particles may either come from the accretion disk and thus be of hadronic origin, or due to pair creation via annihilation processes close to the black hole horizon. We assume that the particle number is conserved,
\begin{align}
& \nabla_{a}(n u^{a})=0 \label{tmasse}\,.
\end{align}
The jet dynamics is governed by the momentum equation,
\begin{align}
& \nabla_{b} T^{ab}=0 \label{tforce}\,.
\end{align}
The first law of thermodynamics is obtained by projecting the conservation of the energy-momentum tensor along the fluid 4-velocity,
\begin{align}
u_{a}\nabla_{b} T^{ab}=0\,. \label{first_principle}
\end{align}

The basic steady equations governing the kinematics of the outflow (\ref{tmasse}) and (\ref{tforce}) are in the 3+1 form,
 \begin{align}
\vec{\nabla}.(h \gamma n \vec{V})=0 \label{masse}\,,\\
 \gamma n (\vec{V}.\vec{\nabla})\left(\frac{\gamma w\vec{V}}{c^2}\right)
= -\gamma^2 n w \left(\vec{\nabla}\ln h_{} + \frac{\varpi \omega V_{\varphi}}{h\,c^{2}}\vec{\nabla} \ln \omega \right)%\nonumber\\
-\vec{\nabla} P + \hat{q}\vec{E}+ \frac{\vec{j}}{c} \times \vec{B} \label{euler}\,.
 \end{align}

Assuming infinite conductivity, the contribution of the electromagnetic field is null in Eq.(\ref{first_principle}). The first law of 
thermodynamics becomes
\begin{align}
n\vec{V}.\vec{\nabla}w=\vec{V}.\vec{\nabla}P\label{thermo}\,.
\end{align}

\subsection{Constants of motion}\label{constants}

Because of flux-freezing, in steady axisymmetric outflows, streamlines and magnetic fieldlines are roped on the same flux tubes of constant 
mass and magnetic flux.  The poloidal components of the velocity and magnetic field can be derived from a stream function $\Psi(r,
\theta)$ and a magnetic flux function $A(r,\theta)$,
\begin{align}
& \vec{V}_{p}= \frac{1}{4 \pi h \gamma n}\frac{\vec{\nabla} \Psi}{\varpi}  \times \vec{e}_{\hat{\varphi}} \,,\\
& \vec{B}_{p}= \frac{\vec{\nabla} A}{\varpi} \times \vec{e}_{\hat{\varphi}} \label{Bp}\,.
\end{align}

The flow along any magnetic flux tube is given in terms of four constants of motion,
the particle flux per unit magnetic flux,
\begin{align}
\Psi_{A}=\frac{\textrm{d}\Psi}{\textrm{d}A}=\frac{4 \pi h \gamma n V_{p}}{B_{p}}\,,
\end{align}
the angular velocity of the fieldlines,
\begin{align}
\Omega = h \frac{V_{\varphi}}{\varpi} - \frac{\Psi_{A}}{4\pi \gamma n}\frac{B_{\varphi}}{\varpi}+\omega \,,\label{isok}
\end{align}
the total angular momentum,
\begin{align}
{\cal L}=\varpi \left(\frac{\gamma w }{c^{2}}V_{\varphi} - \frac{h}{\Psi_{A}}B_{\varphi}\right)\,,\label{angular}
\end{align}
the total energy,
\begin{align}
{\cal E}=h \gamma w - h \frac{\varpi \Omega}{\Psi_{A}} B_{\varphi} + \frac{\gamma w \varpi \omega}{c^{2}} V_{\varphi}\,.\label
{bernoulli}
\end{align}

From Eq. (\ref{isok}) we deduce the bulk velocity of a fluid particle,
\begin{align}
\vec{V}=\frac{\Psi_{A}}{4 \pi h \gamma n} \vec{B} + \frac{\varpi (\Omega - \omega)}{h} \vec{e}_{\varphi} \label{vit}\,.
\end{align}

\subsection{Alfv\'en surface}

By combining Eqs. (\ref{angular}-\ref{bernoulli}) we obtain:
\begin{equation}
  {\cal E} - {\cal L}\Omega = {\cal E} (1 - x_{\rm L}^2)
    = h\gamma w \left[1 - \frac{xV_\varphi}{c}
    \left(1 - \frac{\omega}{\Omega}\right)\right]\,,
  \label{EmLO}
\end{equation}
with
\begin{equation}
  x_{\rm L}^2 = \frac{{\cal L}\Omega}{\cal E}\quad\mathrm{and}\quad
    x = \frac{\Omega\varpi}{ch}\,.
  \label{xL and x}
\end{equation}
Finally, we deduce the values of the toroidal velocity
\begin{equation}
  V_\varphi = \frac{c}{x}\frac{M^2x_{\rm L}^2
    - (1-x_{\rm L}^2)h^2x^2(1-\frac{\omega}{\Omega})}
   {M^2(1-x_{\rm L}^2\frac{\omega}{\Omega})-h^2(1-x_{\rm L}^2)}\,,
\label{V_phi}
\end{equation}
and the relativistic enthalpy
\begin{equation}
  h\gamma w= {\cal E} \frac{M^2(1 - x_{\rm L}^2\frac{\omega}{\Omega})
    - h^2(1-x_{\rm L}^2)}{M^2 - h^2 + h^2x^2(1-\frac{\omega}{\Omega})^2}\,,
\label{Enthalpy}
\end{equation}
where $M$ is the poloidal Alfv\'enic Mach number as defined in \citep{Meliani06}, see also \citep{Michel69,Camenzind86b,BreitmoserCamenzind00},
\begin{align}
M^2=\frac{\Psi_{A}^2 w}{4\pi n c^2} \label{Mach}\,.
\end{align}

At the critical point where the denominators of Eq. (\ref{V_phi}) and Eq.(\ref{Enthalpy}) vanish, the Alfv\'enic Mach number takes the 
value,
 \begin{equation}
 M^2_a= h^2_a\frac{1-x_L^2}{1-x_L^2\frac{\omega_\star}{\Omega}}\,.
\label{M2a}
\end{equation}

It is easily to deduce from the previous equations that  at this point, 

\begin{equation}
 x^2_{\rm a}=   x_{\rm L}^2\frac{1}{(1-\frac{\omega_\star}{\Omega})(1-x_L^2\frac{\omega_\star}{\Omega})}\,.
\label{xa}
\end{equation}

\subsection{Light surfaces}

The light cylinder is defined as the surface where $\|\vec{E}_{p}\|=\|\vec{B}_{p}\|$. From Eqs. (\ref{ohm}) and (\ref{isok}) we obtain the poloidal electric field 
$\vec{E}_{p}=-x \vec{B}_{p}$. The position of the two light surfaces is thus given by $x=\pm1$. 

In the following, we assume that we can neglect the effect of the electric field compared to the magnetic field, ${E}_{p}<<{B}_{p}$ therefore we shall consider $x\rightarrow0$ in Eqs. (\ref{V_phi}-\ref{Enthalpy}). The Alfv\'en regularity condition rewrites 
 \begin{equation}
 M_a^2= h_a^2
 \label{regular}
\end{equation}
As a consequences, the field lines cannot cross the light cylinder, in contrast with radially self similar disk wind solutions  \citep{VlahakisKonigl03a, 
VlahakisKonigl03b}. We are limited to describing jets possessing a weak rotation velocity compared to the speed of light; our solutions are pressure 
driven. The typical radius of the spine jet emerging of the black hole is a few dozen Schwarzschild radii.

\section{Model functions and parameters}\label{model}

In this section, we present the mathematical formalism of the $\theta-$self-similar model. According to the notations used in \citep{ST94} all physical quantities are normalized in units of the Alfv\'en quantities at the pole (subscript $\star$). We have assumed as in previous papers that all physical quantities can be developed to first order in $\alpha$, which is equivalent to assuming small colatitudes. The model hypotheses lead to 4 normalized functions from which two control the flow geometry,
\begin{itemize}
\item
The function $G(R)$ gives the variations of the cylindrical radius $\varpi$ with the distance to the origin, $R$.
\item
The function $F(R)$ is the expansion factor. It measures the angle between the poloidal magnetic field  $\vec{B}_P$ and the radial direction $\vec{e}_R$ \cite{ST94,Meliani06}.
\end{itemize}
Two functions control the flow dynamics, 
\begin{itemize}
\item
The first is the poloidal Alfv\'en Mach number $M^2(R)$;
\item
The second is $\Pi(R)$ which measures the pressure along the polar axis. \\
\end{itemize}

The model possesses six free parameters,
\begin{itemize}
\item
$\delta$ is the latitudinal variation of density. For $\delta > 0$ ($\delta < 0$) density increases (decreases) going out from the axis.
\item
$\kappa $ is the latitudinal variation of the pressure. For $\kappa > 0$ ($\kappa < 0$) pressure increases (decreases) going out from the axis.
\item
$\lambda$ measures the rotation of the flow at the Alfv\'en surface. It is also a measure of the magnetic lever arm.
\item
$\nu$ is the ratio of the escape velocity in units of the Alfv\'en speed at the Alfv\'en surface. 
\item
$\mu$ is the normalized Schwarszchild radius.
\item
$\sigma$ measures the black hole spin. For the maximal rotation value $a/m=1$, $\sigma=\mu/2$.
\end{itemize}

\subsection{Gravitational potential}\label{gravity}
The normalized spherical radius and the normalized spin parameter are, respectively,
\begin{align}
&R=\frac{r}{r_\star}\,,
&\sigma=\frac{a}{r_\star}\,.
\end{align}

The strength of the gravitational potential is given by the polar escape speed at the 
Alfv\'en point in units of $V_\star$,
\begin{align}
\nu=\frac{V_{{\rm esc,}\star}}{V_\star}=\sqrt{\frac{2m}{r_{\star}}}\frac{c}{V_{\star}}\label{Defnu}\,.
\end{align}
The normalized Schwarzschild radius introduced by \citep{Meliani06} is given by the parameter $\mu$,
\begin{align}
\mu=\frac{r_{\rm S}}{r_\star}=\frac{2m}{r_{\star}}\label{Defmu}\,,
\end{align}
which is also the escape speed in units of the speed of light.
Combining Eqs. (\ref{Defnu}) and (\ref{Defmu}) we get a condition that
restrict the parametric space to
\begin{align}
\frac{\sqrt{\mu}}{\nu}=\frac{V_\star}{c}<1\,.
\end{align}
To recover the $\theta$-self-similar model \citep{ST94}, we must expand the metric to first order in $\sin^2\theta$. 
The redshift factor and the frame-dragging angular velocity write,
\begin{align}
h\equiv\sqrt{1-\frac{\mu R}{R^2+\sigma^2}}\left(1-\frac{\mu\sigma^2R}{2(R^2+\sigma^2)^2}\sin^2\theta\right)\,,
\end{align}
\begin{align}
\omega\equiv\frac{\mu\sigma Rc}{r_\star(R^2+\sigma^2)^2}\left[1+\frac{\sigma^2}{R^2+\sigma^2}\left(1-\frac{\mu R}{R^2+
\sigma^2}\right)\sin^2\theta\right]\,.
\end{align}

\subsection{Magnetic field geometry}

Let us introduce a normalized magnetic flux function $\alpha$.  The magnetic flux potential can be separated into a function of  $R$ times a function of $\theta$. To first order in $\alpha$, keeping the dipolar term,  we have,
\begin{align}
\alpha=f(R)\sin^2 \theta
\end{align}

As pointed out in Sect.\ref{model-des}, in meridionally self-similar flows, the $\theta-$dependance is prescribed and the Bernoulli and Grad-Shafranov equations are solved for $f(R)$. 
Hence $f(R)$ measures the magnetic flux relative derivative and its profile determines the geometry of the streamlines. 
This function contains the normalized cylindrical radius $G(R)$ and is related to the expansion factor $F(R)$ as described below.

We get for the poloidal magnetic field from Eq.  (\ref{Bp}),
\begin{equation}
\vec{B_{p}}= 
  \left(
      \begin{aligned}
       \frac{r_{\star}^2 B_\star}{\Sigma}f\cos\theta\\
       -\frac{r_{\star}B_{\star}\sqrt{\Delta}}{2 \Sigma}\frac{df}{dR}\sin\theta
      \end{aligned}
    \right)
\end{equation}

which gives keeping only first order terms, 
\begin{equation}
\vec{B_{p}}= 
  \left(
      \begin{aligned}
       \frac{B_\star}{R^2+\sigma^2}f\cos\theta\\
       -\frac{h_{}B_{\star}}{2 \sqrt{R^2+\sigma^2}}\frac{df}{dR}\sin\theta  
      \end{aligned}
    \right)
\end{equation}

The form of the magnetosphere is determined by the explicit dependence of the magnetic flux function 
$\alpha(R,\theta)$ on its variables. Using Stokes's theorem, Eq.(\ref{Bp}) leads to $2\pi A = \oint_{S}\vec{B}_{p}.\vec{dS}=\pi r_\star^2 B_\star \alpha$,
where
\begin{align}
\varpi_a=r_\star\sqrt{\alpha}\label{alf_radius}
\end{align}
is the \textit{cylindrical} radius at the Alfv\'en point, in other words the magnetic lever arm.
We introduce the function $G(R)$ which represents the normalized cylindrical radius of a given flux tube,  
\begin{align}
G(R)=\frac{\varpi}{\varpi_a}\,. 
\end{align}
We assume that $\alpha$ has a dipolar latitudinal dependence $\alpha\propto\sin^2\theta$ that is consistent with the Taylor development of the cylindrical 
radius, 
\begin{align}
\varpi \equiv r_\star \sqrt{R^2+\sigma^2}\sin\theta\label{cyl_radius}\,. 
\end{align}
Combining Eqs.(\ref{alf_radius}-\ref{cyl_radius}) we obtain the magnetic flux function, 
\begin{align}
\alpha=\frac{R^2+\sigma^2}{G^2}\sin^2\theta\,. 
\end{align}

We introduce the expansion factor $F(R)$ that is the logarithmic derivative of $\alpha$ with respect to the radius $R$,
$F=\left. {\partial \ln{\alpha}}/{\partial \ln{R}}\right|_\theta
\label{F-G} $.
$F(R)$ reflects the degree of the flow collimation (see \citep{ST94, STT99} for more details).

\subsection{Density and pressure functions}
The crucial assumption relates to the shape of the Alfv\'en surface. 
In the absence of an accretion disk the shape of this surface emerges as ellipsoidal in the numerical modeling of \citet{Sakurai85}.
In our model, we assume that this critical surface is spherical, $M(R,\alpha)\equiv M(R)$. The Alfv\'en regularity condition becomes  $M_{\star}= h_{\star}$.
Using Eq.(\ref{Mach}) we find the equipartition relation between electromagnetic and kinetic energy,
\begin{align}
\frac{B_{\star}^2}{8\pi} = \frac{1}{2} n_{\star}w_{\star}\gamma_{\star}^2\frac{V_{\star}^2}{c^2}\,.
\end{align}

The assumption of a spherical Alfv\'en surface implies a separation of the variables in the expression of the free function $\Psi_A$ (see Eq.\ref{Mach}). Making a first order expansion in $\alpha$, 
\begin{align}
\Psi_A^2=4\pi c^2 h_\star^2\frac{n_\star}{w_\star}(1+\delta\alpha)\label{PSIA}\,,
\end{align}
where $\delta$ is a free parameter describing the deviation from spherical symmetry of the ratio $n/w$. Conversely to the classical model it is not the deviation of the mass density itself as in \citet{ST94}). 
Following Eq.(\ref{thermo}) the $\theta$-dependance of the gas pressure is similar to that of the density to enthalpy ratio, 
\begin{align}
P=P_0 + \frac{B_\star^2}{8\pi} \Pi(R)(1+\kappa\alpha) \label{pressure}
\end{align}
with $P_0$  a constant, $\kappa$ a free parameter describing the deviation from spherical symmetric pressure, and $\Pi(R)$ a normalized function.

\subsubsection{Angular momentum flux and the Bernoulli constant}
The function ${\cal L}\Psi_A$ expresses both the total angular momentum loss rate per unit of magnetic flux and the poloidal electric current carried by the outflow \citep{Meliani06}.

\begin{align}
{\cal L}\Psi_A=\lambda h_{\star} B_{\star}r_{\star} \alpha \label{LPSI}
\end{align}
Using Eq.(\ref{PSIA}) we expand the angular momentum ${\cal L}$ to first order,
\begin{align}
{\cal L}=\lambda \gamma_\star w_\star  r_\star \frac{V_\star}{c^2} \alpha \label{L}
\end{align}
The functions ${\cal L}$ and $\Omega$ are free while their ratio is fixed by the regularity condition (\ref{regular}). 
We deduce from Eq.(\ref{regular}-\ref{L}) the Bernoulli integral and the isorotation law, 
\begin{align}
&{\cal E}=h_{\star}\gamma_{\star}w_\star\,,
&\Omega-\omega=(\lambda - \bar{\omega})\,\frac{h_{\star}V_{\star}}{r_{\star}}\,,
\end{align}
As in  \citep{Meliani06} the $\lambda$ parameter measures the strength of the magnetic torque. In the following we introduce a new parameter $\aleph=1-{\bar{\omega}}/{\lambda}$ to introduce the normalized frame-dragging potential $\bar{\omega}$ in our equations.

\subsection{Expressions for the fields and the enthalpy}

From the previous assumptions we get the expressions of the magnetic and velocity fields as well as the density and the enthalpy functions:
\begin{eqnarray}
\label{Br}
B_r&=&\frac{B_\star}{G^2}\cos{\theta}\,,\\
\label{BQ}
B_{\theta}&=&- \frac{B_\star}{G^2}\frac{h_{}F}{2}\sin{\theta}\,,\\
\label{Bphi}
B_{\varphi}&=&- \lambda\aleph\frac{B_\star}{G}\frac{h_{\star}}{h}
\frac{N_B}{D} \sqrt{\alpha} \,,\\
\label{Vr}
V_r&=&\frac{V_\star M^2}{G^2 h_{\star}^2}
\cos{\theta} \,,\\
\label{VQ}
V_{\theta}&=&-\frac{V_\star M^2}{G^2 h_{\star}^2}
\frac{h_{}F}{2} \sin{\theta}
\,,\\
\label{Vphi}
V_{\varphi}&=&-  \lambda\aleph\frac{V_\star}{G} \frac{h}{h_{\star}}
\frac{N_V}{D}
\sqrt{\alpha}
\,,\\
h_{}\gamma w&=&h_{\star} \gamma_\star w_\star
\left[1-\frac{\mu \lambda^2}{\nu^2}
\left(\aleph^2\frac{N_B}{D}+\frac{\bar{\omega}}{\lambda}\right) \alpha \right] \label{hgw}
\,,\\
 h_{}\gamma n&=&h_{\star} \gamma_\star  n_\star
\frac{h_{\star}^2}{M^2}\left[1+\delta\alpha -\frac{\mu
\lambda^2}{\nu^2} \left(\aleph^2\frac{N_B}{D}+\frac{\bar{\omega}}{\lambda}\right) \alpha \label{hgn}
\right] \,.\end{eqnarray} 
where
\begin{align}
N_B=\frac{h^2 }{h_{\star}^2} - G^2\,,\\
N_V=\frac{M^2}{h_{\star}^2}-G^2\,,\\
D=\frac{h_{}^2 }{h_{\star}^2}-\frac{M^2}{h_{\star}^2}
\,.\end{align}

\section{Results}\label{results}             
             
\subsection{Jet morphology and Lorentz factor}
\begin{figure}
{\rotatebox{0}{\includegraphics[scale=0.2]{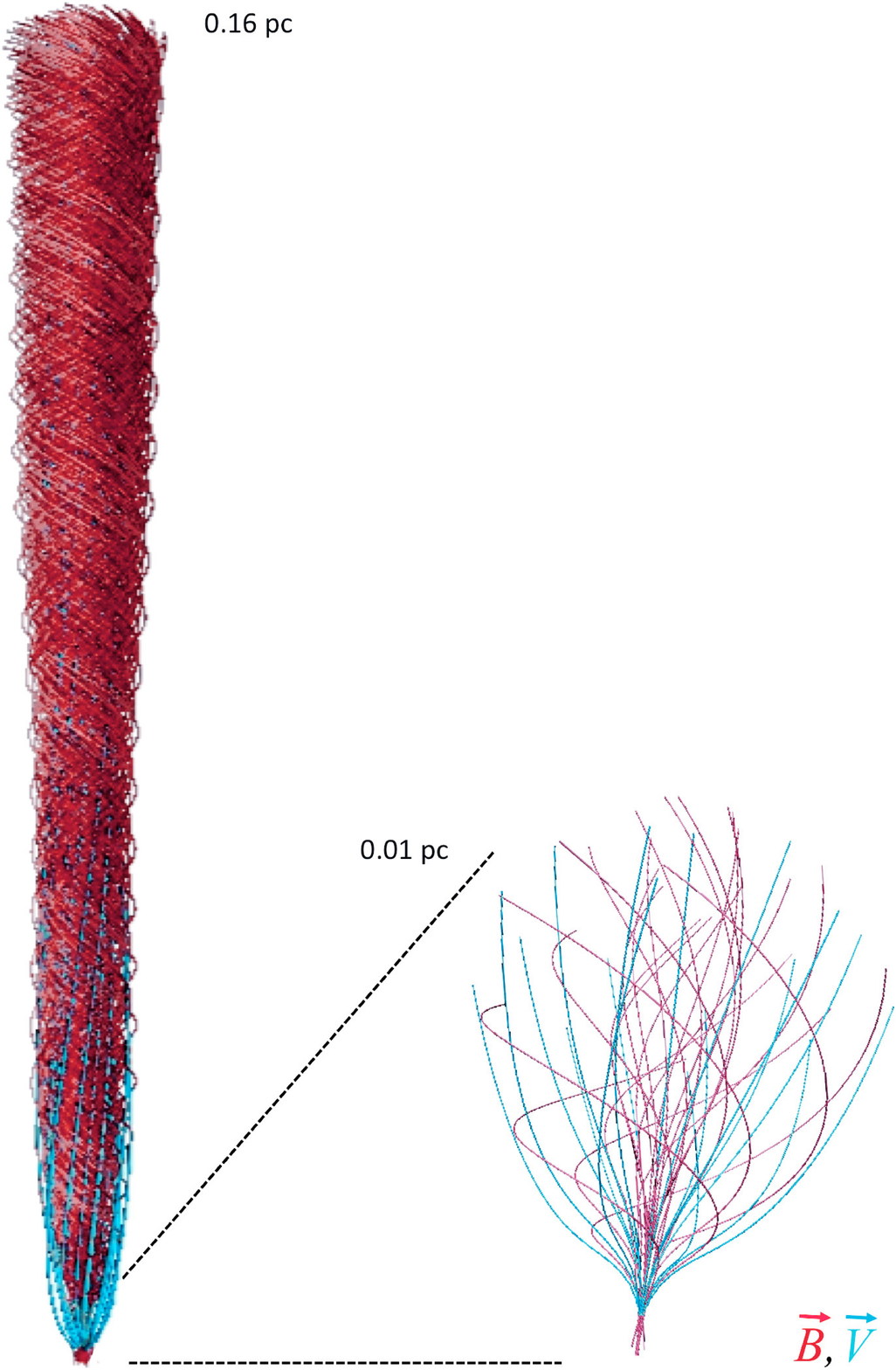}}}
\caption{Spine jet -- topology of the velocity and magnetic field lines of our analytical and stationnary jet solution for a black hole spin parameter $a/m=0.4$.}\label{f2}
{\rotatebox{0}{\includegraphics[scale=0.45]{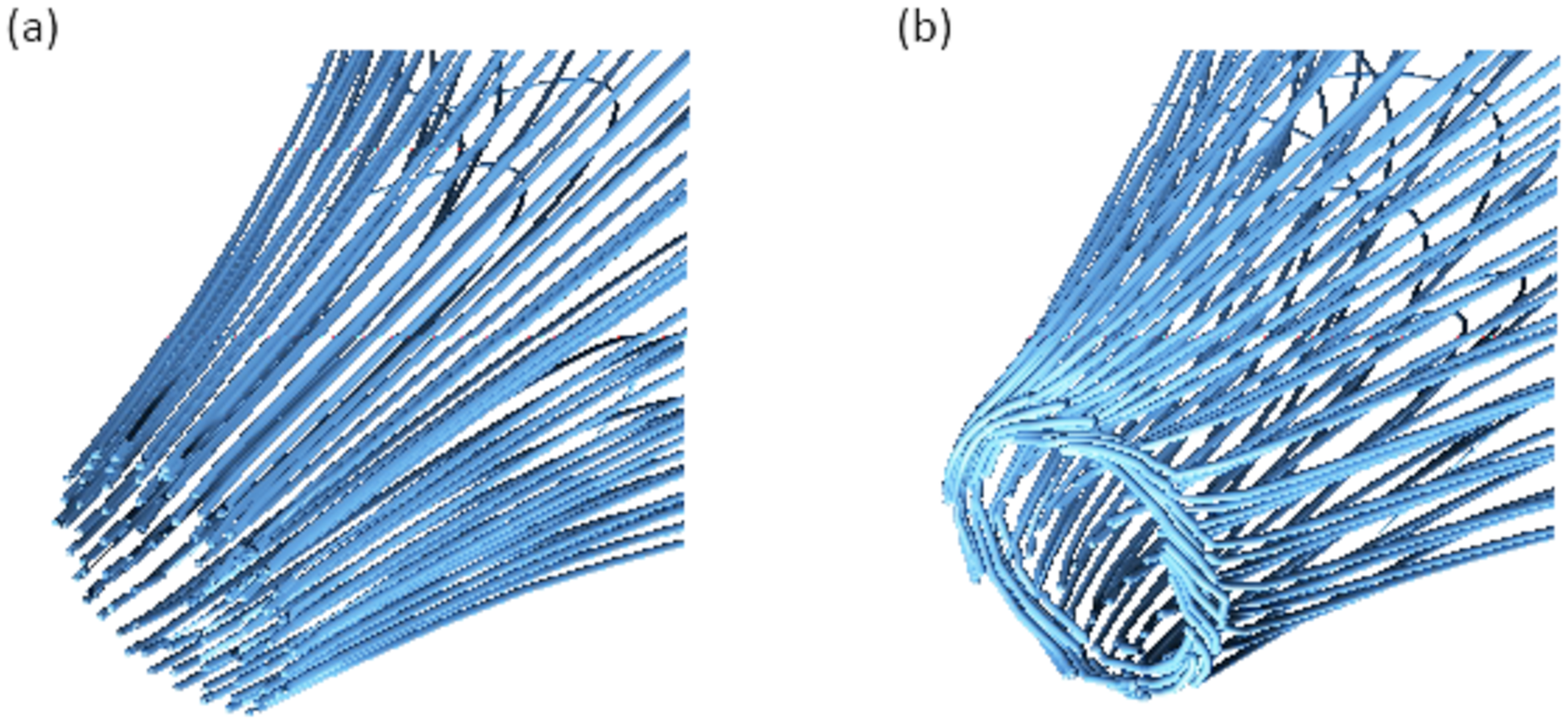}}}
\caption{Spine jet -- topology of the stream lines at the base of the jet in (a) Schwarzschild geometry and (b) Kerr geometry.}
\label{f3}
\end{figure}

\begin{figure}
{\rotatebox{0}{\includegraphics[scale=0.55]{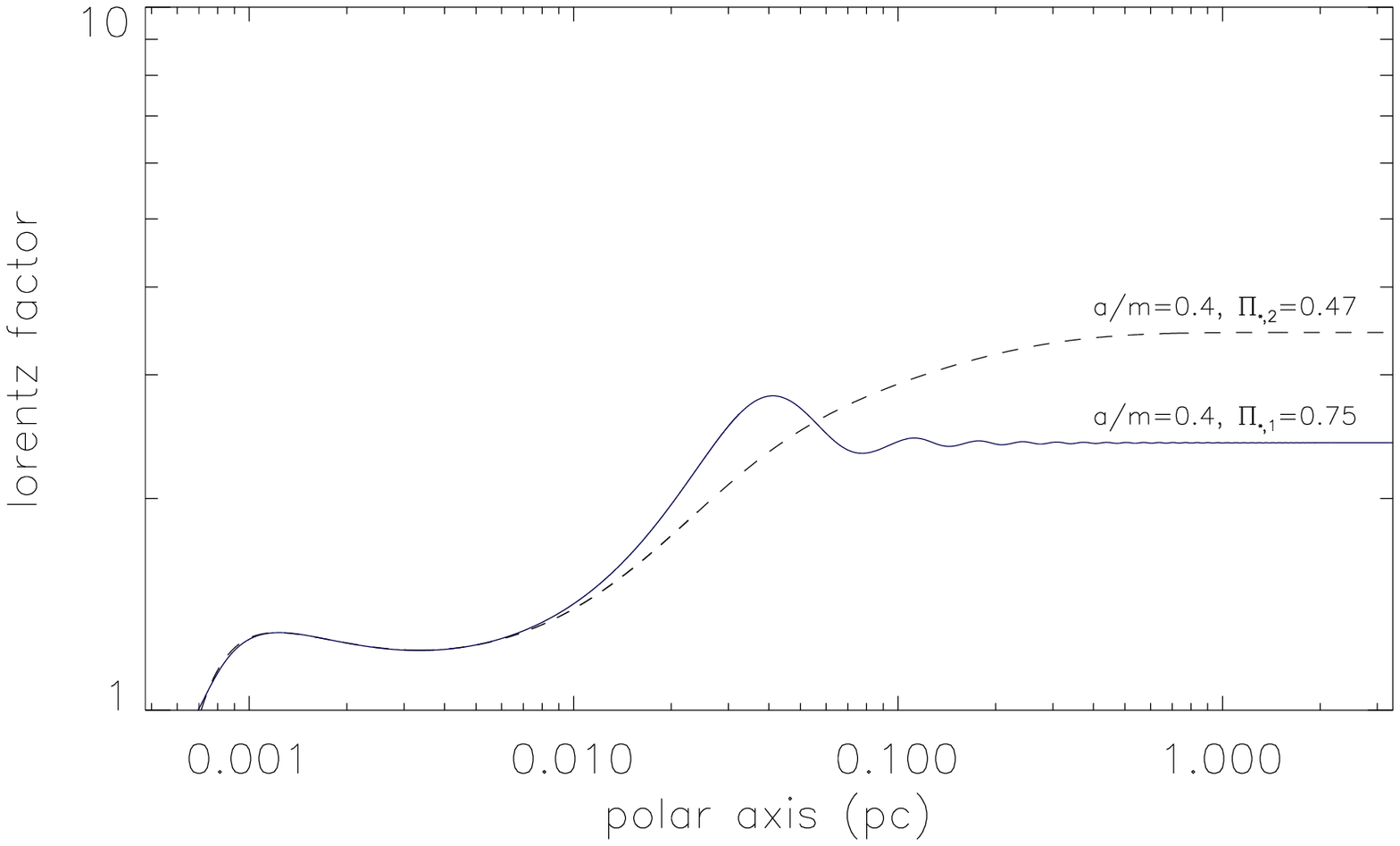}}}
\caption{Lorentz factor profile of the solution displayed in Fig.2. }
\label{f4}
\end{figure}

We solve equations (\ref{ODE1}-\ref{ODE2}). The numerical procedure is given in the appendix. The jet is launched from a region very close to the black hole, typically 2 gravitational radii.
We present solutions corresponding to the following set of parameters :  $\nu=0.54$, $\mu=0.1$, $\lambda=1.1$, $\kappa=0.2$, $\delta=1.35$, and for a black hole spin $a=0.4$. Fig.\ref{f2} shows the topology of the velocity and magnetic field. Fig.\ref{f3} shows the effect of the black hole rotation on the jet base.
Fig.\ref{f4} displays the corresponding Lorentz factor. The asymptotic Lorentz factor is larger in the case of a limiting solution, i.e. when the pressure $\Pi_\star$ is the lowest one that gives a cylindrical solution.

\subsection{Magnetic collimation efficiency}  
\begin{figure}
{\rotatebox{0}{\includegraphics[scale=0.45]{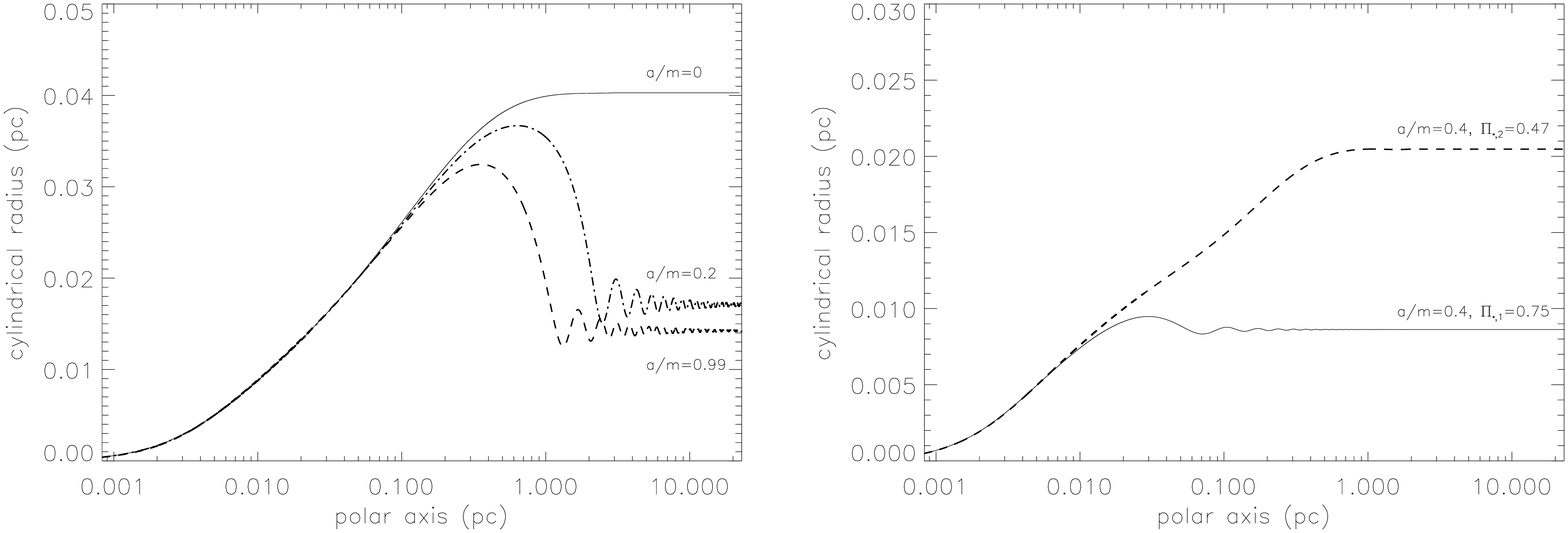}}}
\caption{Dependence of the cylindrical radius on the black hole spin (left panel) and on the pressure (right panel). }
\label{cylradius}
\end{figure}

Fig.\ref{cylradius} (left panel) displays three solutions to the GRMHD equations, corresponding to the same physical parameters, except for the spin of the central engine. ($\nu=1.5$, $\mu=0.1$, $\lambda=0.73$, $\kappa=0.021$, $\delta=0.0778$). The upper jet is launched by a Schwarzschild black hole while the other two solutions by a Kerr black hole ($a/m=0.2$, $a/m=0.99$, respectively). The jet solution in the Schwarzschild case is asymptotically cylindrical because we have chosen the minimum value of $\Pi_\star=\Pi_{\star,1}$ giving the limiting solution (see \citet{STT99, Meliani06} for details). For a lower initial pressure the jet would decollimate while with a higher initial pressure the jet would undergo recollimation with oscillations. An oscillatory jet width is a basic feature of such recollimating jets \citep{STT99}. Oscillations occurs because of the interplay of the centrifugal force and the total (pinching plus pressure) magnetic force, always acting in opposite directions along the cylindrical 
radius $\varpi$. The transverse pressure gradient remaining similar this proves that the rotation of the central black hole enhances the efficiency of the magnetic field to collimate.

We can show in a different manner the same result now using the Kerr solution of Fig. \ref{f2} and lowering the initial value of the pressure by lowering $\Pi_\star$ until we get the limiting solution for a value $\Pi_\star=\Pi_{\star,2}$. Fig. \ref{cylradius} (right panel)  displays two jet morphologies which are almost identical but clearly the same rate of collimation is obtained in the second case for a lower pressure as $\Pi_{\star,2}<\Pi_{\star,1}$.

\subsection{Magnetic collimation parameter}  
In \citet{STT99}, a general criterion for the jet collimation has been established, based on the variation of the energy across poloidal fieldlines. It gives an important extra parameter $\epsilon$ which provides the efficiency of the magnetic rotator to collimate the flow. We follow the same procedure as in \citet{Meliani06} to derive the magnetic collimation parameter $\epsilon$ in the Kerr metric.

After substituting $n$ from Eq. (\ref{Mach}) and derivating with $\alpha$ constant, ${\vec{V}}
\cdot \nabla  \propto {\partial }/{\partial R}|_\alpha$
, Eq. (\ref{thermo})
can be re-written as
\begin{eqnarray}
-8 \pi M^2 \left. \frac{\partial P}{\partial R} \right|_\alpha
&=&-\frac{\partial}{\partial R} \left.\left(
\frac{\Psi_A^2 w^2}{c^2}\right)\right|_\alpha
\nonumber\\
&=&\frac{\partial}{\partial R} \left.\left[
\frac{\Psi_A^2 ({\cal E}^2 - w^2)}{c^2}\right]
\right|_\alpha
\label{integral1}
\,,
\end{eqnarray}
where $\Psi_A^2 w^2$ is proportional to
the energy {\it per unit volume} of the fluid reduced to the thermal content. Thus $\Psi_A^2 ({\cal E}^2-w^2)$ in essence measures the variation between the total energy and the thermal energy of the fluid.

The form of the pressure is $P=f_1(R) (1+\kappa \alpha)/8 \pi$. We also know the $\theta$
dependence in all quantities in the expression for $\Psi_A^2
w^2/c^2$, and after expanding with respect to $\sin^2\theta$ we
find $\Psi_A^2 ({\cal E}^2-w^2)/c^2=f_2(R) + f_3(R) \alpha$. Then
Eq. (\ref{integral1}) gives
\begin{eqnarray}
-M^2 \frac{{\rm d} f_1}{{\rm d}R}(1+\kappa \alpha)
= \frac{{\rm d} f_2}{{\rm d}R} + \frac{{\rm d} f_3}{{\rm d}R} \alpha\nonumber\\
\Leftrightarrow \left\{
\begin{array}{ll}
-M^2 {\rm d} f_1 = {\rm d} f_2 \\
-M^2 \kappa {\rm d}f_1 = {\rm d}f_3
\end{array}
\right.
\end{eqnarray}
Eliminating ${\rm d} f_1$ we get the integral $ f_3(R) - \kappa f_2(R)
= \epsilon$ which is a local measurement of the magnetic efficiency to collimate the flow but does not remain constant along the flow.

After substituting the expressions for $f_2(R)$ and $f_3(R)$, we arrive at
\begin{eqnarray}\label{varepsilon_full}
\epsilon =
\frac{M^{4}}{h_{\star}^4 (R^2+\sigma^2) G^2}
\left(\frac{F^2}{4} - \frac{1}{h_{}^2} - \kappa \frac{R^2+\sigma^2}{h_{}^2G^2}\right)
-\frac{\left(\delta-\kappa\right)\nu^2}{h_{}^2}\frac{R}{R^2+\sigma^2}
\nonumber \\
-\frac{\nu^2\sigma^2RG^2}{h^2(R^2+\sigma^2)^3}
+\frac{\lambda^2}{G^2 h_{\star}^2} \left(\frac{N_V}{D}\right)^2
+\frac{2 \lambda^2}{h_{}^2}\left(\aleph^2\frac{N_B}{D}+\frac{\bar{\omega}}{\lambda}\right)
\,,
\end{eqnarray}

We refer to \citet{STT99} for a detailed parametric analysis and a discussion on the different solutions which are inferred from the modeling. Here we only consider for a first approach $\kappa>0$. This does not necessarily implies that the flow is pressure confined. It may be either magnetically confined or pressure confined depending on whether the efficiency of the magnetic rotator prevails or not to the thermal confinement. A negative $\epsilon$ implies that the source is an inefficient magnetic rotator, which needs the help of the gas pressure to collimate the outflow \citep{STT99, Meliani06}. Conversely, a positive $\epsilon$ is the sign of an efficient magnetic rotator with a strong magnetic contribution to collimation. 
 
It is simple to express this constant at the base of the flow $R_o$ assuming the poloidal velocity is negligible there, such that the Alfv\'enic Mach number $M(R_o)\approx 0$. Then $\epsilon$ takes the following form at the source boundary,
\begin{equation}
\frac{\epsilon}{2\lambda^2}  = \frac{ 
{\cal E}_{{\rm Rot},o}
+ {\cal E}_{{\rm Poynt.},o}
+\Delta {\cal E}_{\rm G}^*+ {\cal E}_{\rm drag}}
{{\cal E}_{\rm MR} }\label{epsilon_source}
\,,
\end{equation}
where ${\cal E}_{\rm Poynt.}=- h_{} \varpi {(\Omega-\omega_o)}/{\Psi_A} B_\varphi$
is the Poynting flux and ${\cal E}_{{\rm R},o}={\cal E}{V_{\varphi, o}^2}/{2c^2}$
is the rotational energy per particle.
The following term is very similar to the nonrelativistic one except for the metric,
\begin{equation}
\Delta {\cal E}_{\rm G}^*
= -\frac{{\cal E}\mu}{2}\frac{R_o}{R_o^2+\sigma^2}\left(\delta-\kappa+ \frac{\sigma^2G_o^2}{(R_o^2+\sigma^2)^2}\right) \alpha\,.
\end{equation}
It measures the excess or the deficit of gravitational energy per unit mass which is not
compensated by the thermal driving \citep{STT99}.
The last term ${\cal E}_{\rm drag}={\cal L}\omega_o$ represents the coupling between the
orbital angular momentum of the fluid particle and the frame dragging of the Kerr black hole.

We see in Eq. (\ref{epsilon_source}) that the magnetic collimation parameter seems to be larger in the Kerr metric, because of the new term ${\cal E}_{\rm 
drag}$ and also because $\Delta {\cal E}_{\rm G}^*$ is smaller as we increase $\sigma$. In order to check this assumption, we calculated solutions with 
the same physical parameters, except for the black hole spin.

\begin{figure}
{\rotatebox{0}{\includegraphics[scale=0.45]{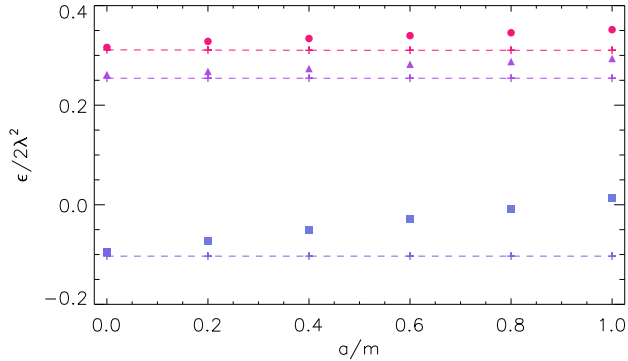}}}
\caption{Dependence of the magnetic collimation efficiency on the black hole spin for two different sets of parameters. The lower solutions (blue squares) 
correspond to: $\nu=1.5$, $\mu=0.1$, $\lambda=0.73$, $\kappa=0.021$, $\delta=0.0778$; the upper solutions (triangles and dots) to : $\nu=0.54$, $
\mu=0.1$, $\lambda=1.1$, $\kappa=0.2$, $\delta=1.35$. In both cases, the magnetic collimation $\epsilon$ increases linearly with the parameter $\sigma$ 
that measures the rotation of the black hole.}
\label{f6}
\end{figure}

Fig.\ref{f6} shows the evolution of $\epsilon$ with the rotation of the black hole for two different sets of parameters corresponding to the solutions 
presented above. The magnetic collimation efficiency increases linearly with the black hole spin. We may expect that it comes with the linear increase of 
the roping of the magnetic field at the base of the flow due to the drift of the rotation coordinate. 

\section{Discussion}\label{discussion}
We extended the meridional self-similar model of \citet{ST94} to the GRMHD case in the Kerr spacetime. We presented the first axisymmetric analytical MHD solutions (2.5D)  for an outflow accelerated near the polar axis of a rotating black hole. In order to keep the self-similarity in colatitude, we had to assume an expansion of all equations including the metric with the colatitude $\theta$. 
As in the self-similar model in the Schwarzschild geometry \cite{Meliani06}, we made the assumption of a non relativistic rotational velocity to neglect the effects of the light cylinder on the Alfv\'en surface. 
We derived exact solutions to the GRMHD equations where the collimation derive from a combination of pressure and magnetic forces. 

The contribution of the different mechanisms to the collimation of the outflow in the context of meridional self-similar models was studied in previous papers \cite{STT99,Meliani06}. In this paper, we made an extension of the magnetic collimation 
parameter for a rotating black hole in order to investigate the effect of the black hole spin on the flow collimation. 
The collimation efficiency, given by the variation of the specific energy across streamlines, has five contributions; each one gives the variation 
- in units of the volumetric energy of the magnetic rotator - between any streamline and the polar axis of: the kinetic energy, 
the volumetric gravitational energy, the Poynting flux, the thermal content and the rotational energy associated with the frame-dragging potential. 
The kinetic and magnetic contents depends strongly on the physics of the central engine. In the picture the jet is driven by the Blandford-Znajek mechanism, the injected power goes as $P\propto\sin^2(\theta)$, \textit{i.e.} there is an energy deficit near the polar axis, 
which would have the effect of increasing the magnetic collimation parameter $\epsilon$. The thermal content may depend on plasma injection along the magnetic field lines, and a complete picture should include the disk wind component, which could inhibit the lateral expansion since it is believed to be more dense than the leptonic spine jet. The rotation
of the black hole could also play a role since the location of the innermost stable orbit depends on the spin parameter. 

The efficiency of self-collimation in relativistic jets can be compromised by the existence of the eletric fields, which are negligible far from the light cylinder. We considered only the region close to the polar axis where the effect of the decollimating electric fields are negligible while the frame-dragging play an important role. 
We found that the magnetic collimation efficiency is enhanced by the rotation of the black hole, and increases linearly with the black hole spin in our model. This gain is due (i) to the fact that the volumetric gravitational energy $\Delta {\cal E}_{\rm G}^*$, 
which has a negative contribution to the collimation efficiency, is smaller in the Kerr metric, and (ii) to the presence of the new term ${\cal L}\omega$, associated with the frame-dragging. 
As our model is valid only for small colatitude angles, the contribution of this term is then under-estimated.

\acknowledgements{We are grateful to Christian Fendt for stimulating discussions, and the anonymous referees for useful coments. 3D images are produced by VAPOR (www.vapor.ucar.edu).}

\begin{widetext}

\appendix
\section{Ordinary differential equations}\label{appendixA}
For convinience we use the following notations : $X_{_+}=R^2+\sigma^2$, $X_{_-}=R^2-\sigma^2$. Under the assumptions of axisymmetry and meridional self-similarity, the GRMHD equations reduce to the following four ordinary differential equations for 
$\Pi(R)$, $M^2(R)$, $G(R)$ and $F(R)$:
\begin{eqnarray}\label{ODE}
\frac{{\rm d}  \Pi}{{\rm d} R}\,\, &=&
-\frac{2}{{h_{} }^2}\frac{1}{G^4}
\left(\frac{{\rm d}  M^2}{{\rm d} R}+\frac{F-2}{R}
 M^2\right)
-\frac{1}{h_{}^4 M^2}\frac{X_{_-}}{X_{_+}^2}\left(\nu^2 h_{\star}^4-\mu\frac{M^4}{G^4}\right)\,, \label{ODE1}\\
\frac{{\rm d}  M^2}{{\rm d} R} &=& \frac{{\cal N}_M}{{\cal D}}\,,\\
\frac{{\rm d}  F}{{\rm d} R} \,\,&=& \frac{{\cal N}_F}{{\cal D}}\,,\\
\frac{{\rm d} G}{{\rm d} R} \,\,&=&\frac{G}{R} \left(1-\frac{F}{2}\right).\label{ODE2}
\end{eqnarray}
where we have defined :
%%%%%%%%%%%%%%%%%%%%%%% NM %%%%%%%%%%%%%%%%%%%%%%%%%%%%%%
\begin{eqnarray}\label{dM2dR}
{\cal N}_M &=& \frac{{M}^4}{4h_{\star}^2 R}
\left[(8-4F)\left(1-\frac{\sigma^2}{2X_{_+}}\right)+4\kappa\frac{X_{_+}}{G^2}(2-F) 
+ \frac{2FR}{\sqrt{X_{_+}}}
\left(1+\frac{\sigma^2}{X_{_+}}\right)
-\frac{h_{}^2F^2R^2}{X_{_+}}-2F\mu\frac{R^2\sigma^2}{X_{_+}^{\frac{5}{2}}}\right]
\nonumber\\&&%%%%%%%%%%%%%%%%%%%%%%%%%%%%%%%%%%%%%%%
+\frac{{h_{}^2 M^2}}{h_{\star}^2}
\left[\frac{F-2}{R}\left(1+\kappa\frac{X_{_+}}{G^2}-\frac{\sigma^2}{2 X_{_+}}\right)
-\frac{F}{\sqrt{X_{_+}}}
+\frac{h_{}^2F^2}{4} \left (\frac{R}{X_{_+}}+\frac{\mu}{h_{}^2}\frac{X_{_-}}{X_{_+}^2} \right)  
+\frac{h_{}^2F^3}{8\sqrt{X_{_+}}} -\frac{h_\star^2}{h^2}\frac{\lambda^2\aleph^2\mu}{\nu^2}F\sqrt{X_{_+}}\right]
\nonumber\\&&%%%%%%%%%%%%%%%%%%%%%%%%%%%%%%%%%%%%%%%
-\frac{D G^2}{2{h_{} }^2 M^2}\frac{X_{_-}}{X_{_+}}\left(\nu^2 h_{\star}^4-\mu\frac{M^4}{G^4}\right)\left[\delta-\kappa+\frac{\sigma^2 G^2}{X_{_+}^2}\left(\frac{1}{2}-h^2\frac{\sigma^2-3R^2}{X_{_-}}\right)-\frac{2\mu \lambda^2 \aleph^2}{\nu^2}\left(\frac{N_B}{D}+\frac{\bar{\omega}}{\lambda \aleph}\right)\right]
\nonumber\\&&%%%%%%%%%%%%%%%%%%%%%%%%%%%%%%%%%%%%%%%
-\frac{M^2\mu}{2h_{}^2}\frac{X_{_-}}{X_{_+}^2}D
+\lambda^2 \aleph^2 \frac{X_{_-}}{X_{_+}}\frac{N_B N_V}{D^2}\mu 
+\frac{h^2 G^2 D}{2}\kappa\Pi\left[\frac{h^2}{h_{\star}^2}\frac{F}{D}\sqrt{X_{_+}}-2R-\frac{X_{_+}}{R}(F-2)\right]
\nonumber\\&&%%%%%%%%%%%%%%%%%%%%%%%%%%%%%%%%%%%%%%%
+ \frac{\lambda \aleph\sqrt{\mu}}{\nu}  \frac{\sigma G^2}{h_{\star}M^2}\left(\nu^2 h_{\star}^4-\mu\frac{M^4}{G^4}\right) \frac{\sigma^2-3 R^2}{X_{_+}^2}N_V
+\frac{h_{}^2}{h_{\star}^2}\frac{F M^2}{2} \frac{R\sigma^2}{X_{_+}^{\frac{5}{2}}} \left(\mu-h_{\star}^4\nu^2\frac{G^4}{M^4}\right)
\nonumber\\&&%%%%%%%%%%%%%%%%%%%%%%%%%%%%%%%%%%%%%%%
-\lambda^2 \aleph^2 h^2 \frac{N_B}{D}(F-2) \frac{X_{_+}}{R}  
%+\mu \lambda^2 \aleph^2\frac{G^2 h_{\star}^4 D}{h_{}^2 M^2} \frac{X_{_-}}{X_{_+}}\frac{N_B}{D}
%+\mu \lambda \aleph\frac{G^2 h_{\star}^4 D}{h_{}^2 M^2} \frac{X_{_-}}{X_{_+}}\bar{\omega}
%\nonumber\\&&%%%%%%%%%%%%%%%%%%%%%%%%%%%%%%%%%%%%%%%
+\lambda^2\aleph^2 R\left[2M^2+h^2\left(\frac{F\sqrt{X_{_+}}}{R}-2\right)\right]\left(\frac{N_B^2}{D^2}-\frac{h^2}{2 M^2}\frac{N_V^2}{D^2}\right)
\,,
\end{eqnarray}

%%%%%%%%%%%%%%%%%%%%%%% NF %%%%%%%%%%%%%%%%%%%%%%%%%%%%%%
\begin{eqnarray}\label{dFdR}
{\cal N}_F&=&
-\frac{ M^2F}{h_{\star}^2\sqrt{X_{_+}}}\left[\Upsilon\left(\frac{R}{\sqrt{X_{_+}}}-\frac{F}{2}\right)-\lambda^2\aleph^2\frac{N_B^2}{D^3}X_{_+}(F-2)\frac{\sqrt{X_{_+}}}{R}
+\frac{F}{2}\left(1-\frac{h^2F R}{2\sqrt{ X_{_+}}}\right)\right]
\nonumber\\&&  %%%%%%%%%%%%%%%%%%%%%%%%%%%%%%%%%%%%%%%
+\frac{h_{}^2}{h_{\star}^2}
\left(\Upsilon-\frac{h_{}^2F^2}{4}\right)
\left[\frac{F^2}{R}+F\left(\frac{R}{X_{_+}}-\frac{2}{R}\right)-\frac{2}{h_{}^2\sqrt{X_{_+}}}
{- 4\frac{h_\star^2}{{h_{}}^4}\frac{\lambda^2\aleph^2\mu}{\nu^2}\sqrt{X_{_+}}}\right]
\nonumber\\&&%%%%%%%%%%%%%%%%%%%%%%%%%%%%%%%%%%%%%%%%
+\left(\Upsilon-\frac{h_{}^2F^2}{4}\right)\frac{\mu F}{h_{\star}^2}\frac{X_{_-}}{X_{_+}^2} 
+\frac{2\Pi G^2\kappa\sqrt{X_{_+}}}{h_{\star}^2}
\left[ \Upsilon-\frac{h_{}^2F^2}{4}   \left(\frac{2}{F}\left[\frac{R}{\sqrt{X_{_+}}}-\frac{\sqrt{X_{_+}}}{R}\right]+\frac{\sqrt{X_{_+}}}{R}\right)   \right]
\nonumber\\&&%%%%%%%%%%%%%%%%%%%%%%%%%%%%%%%%%%%%%%%%
-\frac{G^2 F}{2 M^2 h_{}^2h_{\star}^2}\frac{X_{_-}}{X_{_+}}\left(\nu^2 h_{\star}^4-\mu\frac{M^4}{G^4}\right)\left[\delta-\kappa+\frac{\sigma^2 G^2}{X_{_+}^2}\left(\frac{1}{2}-h^2\frac{\sigma^2-3R^2}{X_{_-}}\right)-\frac{2\mu \lambda^2 \aleph^2}{\nu^2}\left(\frac{N_B}{D}+\frac{\bar{\omega}}{\lambda \aleph}\right)\right]
\nonumber\\&&%%%%%%%%%%%%%%%%%%%%%%%%%%%%%%%%%
+\frac{4\lambda^2\aleph^2}{h_{}^2}  \left(\frac{N_B^2}{D^2}-\frac{h_{}^2}{2 M^2}
\frac{N_V^2}{D^2}\right)\left(\Upsilon\sqrt{X_{_+}}-\frac{h^2 R F}{2}\right)-\lambda^2\aleph^2 \frac{h^2}{h_\star^2}\frac{X_{_+}}{R}F(F-2)\frac{N_B}{D^2}
-\frac{\mu F M^2}{2 h_{}^2 h_{\star}^2}\frac{X_{_-}}{X_{_+}^2}
\nonumber\\&&%%%%%%%%%%%%%%%%%%%%%%%%%%%%%%%%%%
%+\mu \lambda^2\aleph^2 \frac{G^2 h_{\star}^2 F }{h_{}^2 M^2} \frac{X_{_-}}{X_{_+}}\frac{N_B}{D}
%+\mu \lambda\aleph \frac{G^2 h_{\star}^2 F }{h_{}^2 M^2} \frac{X_{_-}}{X_{_+}}\bar{\omega}
+\frac{2 M^2 \sigma^2}{ h^2 h_{\star}^2 X_{_+}^{\frac{3}{2}}}\left[\left(1-\frac{\nu^2 h_{\star}^4G^4R}{M^4 X_{_+}}\right)
\Upsilon-\frac{h_{}^2F^2}{4}\right]
+\lambda^2\aleph^2\frac{\mu}{h_{\star}^2}\frac{X_{_-}}{X_{_+}}\frac{N_B N_V}{D^3}F
+\mu\frac{F^2 M^2}{2 h_{\star}^2}\frac{R\sigma^2}{X_{_+}^{\frac{5}{2}}}
\nonumber\\&&%%%%%%%%%%%%%%%%%%%%%%%%%%%%%%%%%%
%-\frac{G^4 F \sigma^2}{2 h^2 h_{\star}^2M^2}\frac{X_{_-}}{X_{_+}^3}\left(\nu^2 h_{\star}^4-\mu\frac{M^4}{G^4}\right)\left(\frac{1}{2}-h^2\frac{\sigma^2-3R^2}{X_{_-}}\right)
%\nonumber\\&&%%%%%%%%%%%%%%%%%%%%%%%%%%%%%%%%%%
+ \frac{\lambda \aleph\sqrt{\mu}}{\nu} \frac{G^2 F\sigma}{M^2h_{\star}^3}\left(\nu^2 h_{\star}^4-\mu\frac{M^4}{G^4}\right) \frac{\sigma^2-3 R^2}{X_{_+}^2}\frac{N_V}{D}
\,,
\end{eqnarray}

\begin{eqnarray}
{\cal D} &=& -\left(1-\frac{\sigma^2}{2X_{_+}}+\kappa\frac{X_{_+}}{G^2}\right) D
+\lambda^2 \aleph^2 X_{_+}\frac{N_B^2}{D^2}+\frac{h_{}^4 F^2}{4h_{\star}^2}\,,\\
\Upsilon &=& 1-\frac{\sigma^2}{2X_{_+}}+\kappa\frac{X_{_+}}{G^2}-\lambda^2\aleph^2\frac{N_B^2}{D^3}X_{_+}
\,.\end{eqnarray}

At the Alfv\'en radius, the expansion factor is the solution of the second degree polynomial $C_2 F_{\star}^2+C_1 F_{\star}+ C_0 = 0 $, with 
\begin{eqnarray}
C_0 &=&  \frac{- h^2_\star \sqrt{X_{+,\star}} p^{' 2}}{8} - \frac{\lambda^2 \aleph^2_\star X_{+,\star}^{3/2}}{2} h^4_\star\,,\\
C_1 &=& \frac{X_{+,\star} p^{' 3}}{4} + \lambda^2 \aleph^2_\star X_{+,\star}^{3/2} h^2_\star (p' + 2 h^2_\star -\mu \Lambda_\star)\,,\\
C_2 &=&p^{' 2} \left[ \lambda^2 \aleph^2_\star X_{+,\star}^{3/2} \left( \frac{\mu}{\nu^2} + \frac{1}{2}\right) + \frac{\sqrt{X_{+,\star}}}{2} ( 1- \kappa \,  \Pi_\star X_{+,\star}) - \frac{\sigma^2}{2 \sqrt{X_{+,\star}}} \left( 1 - \frac{\nu^2}{X_{+,\star}}\right) \right] - \lambda^2 \aleph^2_\star X_{+,\star}^{3/2} p' \left[  2 h^2_\star - \mu \Lambda_\star \right] \nonumber \\
&& - \lambda^2 \aleph^2_\star X_{+,\star}^{3/2} \left[ 2 h^4_\star - 2 \mu \Lambda_\star h^2_\star + \frac{\mu^2 \Lambda^2_\star}{2}\right]\,.
\end{eqnarray}
where $X_{_+\star}=1+\sigma^2$, $X_{_-\star}=1-\sigma^2$, $\aleph_\star=1-\frac{\bar{\omega}_\star}{\lambda}$.

\section{Overview of the numerical technique}

Assuming self-similarity, the PDE system of GRMHD equations is transformed into an ODE system where the functions depends only on the normalized radial distance $R$. Four coupled equations constitute the system : the three equations given in appendix \ref{appendixA} determine the unknown functions $\Pi(R)$, $F(R)$ and $ M^2(R)$ and the function $G(R)$ is related to $F(R)$ through Eq. (\ref{F-G}).
We have shown that $F(R)$, $G(R)$ are dimensionless functions related to the shape of the jet while $M(R)$, $\Pi(R)$ describe the physics of the magnetosphere.
We briefly present the method for the numerical integration of Eqs. (\ref{ODE1}) - (\ref{ODE2}).
We start integrating the equations from the Alfv\'en critical
surface. In order to calculate the toroidal components of the fields, i.e.
$N_B/D$ and $N_V/D=N_B/D-1 $,  we apply l'Hospital's rule at this point,
\begin{equation}
\left.\frac{N_B}{D}\right|_\star =\frac{h_{\star}^2(2- F_\star)
-\mu\left[\frac{1-\sigma^2}{(1+\sigma^2)^2}\right] } {p-\mu\frac{1-\sigma^2}{(1+\sigma^2)^2}} \,, \quad
p=\left.\frac{{\rm d}  M^2}{{\rm d} R}
\right|_{\star}\,.
\end{equation}

To avoid kinks in the fieldline shape, we need to satisfy
a regularity condition \cite{HeyvaertsNorman89}. This means that Eq.
(\ref{dFdR}) should be regular at $R=1$. As in \citep{Meliani06} this extra
requirement is equivalent to ${\cal N}_F.D= 0$ which eventually
gives a second order polynomial equation for the Alfv\'enic expansion factor $F_{\star}$,
\begin{equation}
C_2(p) F_{\star}^2+C_1(p) F_{\star}+ C_0(p,\Pi_{\star}) = 0 \,,
\end{equation}
with the expressions for the coefficients $C_2$, $C_1$ and $C_0$ given in the appendix \ref{appendixA}.
Once we have determined the regularity conditions at the Alfv\'en point, we
integrate downwind and upwind and cross all the other existing
critical points as in the non relativistic case.
Notice that solutions depend also on  $\Pi_{\star}$, i.e.  the pressure at the Alfv\'en
surface.  As in the classical case its value has been chosen such that the total gas pressure is always
positive. More details on the numerical technique can be found in  \citep{ST94,STT99, Meliani06}.

\end{widetext}

\end{document}